\RequirePackage{amsmath}
\documentclass{elsarticle} 

\usepackage{graphicx}
\usepackage{subcaption}
 \usepackage{caption}
\usepackage{amsfonts}
\usepackage{booktabs}
\usepackage[numbers]{natbib}
\usepackage{notoccite}
\usepackage{eucal}
\usepackage{verbatim}
\usepackage{physics}

\usepackage{lineno}

\usepackage{color}
\usepackage{tikz}
\usetikzlibrary{shapes}

\usepackage{setspace}
\usepackage{bigints}
\usepackage{siunitx}
\usepackage{lipsum}
 \usepackage{xcolor}
 \usepackage{soul}
 
\usepackage{booktabs}
\usepackage{multirow}
 \usepackage{hhline}

\usepackage[english]{babel}
\usepackage{blindtext}

 \usepackage{booktabs}
    \usepackage{tabularray}
\makeatletter
\let\save@mathaccent\mathaccent
\newcommand*\if@single[3]{%
  \setbox0\hbox{${\mathaccent"0362{#1}}^H$}%
  \setbox2\hbox{${\mathaccent"0362{\kern0pt#1}}^H$}%
  \ifdim\ht0=\ht2 #3\else #2\fi
  }
\newcommand*\rel@kern[1]{\kern#1\dimexpr\macc@kerna}
\newcommand*\widebar[1]{\@ifnextchar^{{\wide@bar{#1}{0}}}{\wide@bar{#1}{1}}}
\newcommand*\wide@bar[2]{\if@single{#1}{\wide@bar@{#1}{#2}{1}}{\wide@bar@{#1}{#2}{2}}}
\newcommand*\wide@bar@[3]{%
  \begingroup
  \def\mathaccent##1##2{%
    \let\mathaccent\save@mathaccent
    \if#32 \let\macc@nucleus\first@char \fi
    \setbox\z@\hbox{$\macc@style{\macc@nucleus}_{}$}%
    \setbox\tw@\hbox{$\macc@style{\macc@nucleus}{}_{}$}%
    \dimen@\wd\tw@
    \advance\dimen@-\wd\z@
    \divide\dimen@ 3
    \@tempdima\wd\tw@
    \advance\@tempdima-\scriptspace
    \divide\@tempdima 10
    \advance\dimen@-\@tempdima
    \ifdim\dimen@>\z@ \dimen@0pt\fi
    \rel@kern{0.6}\kern-\dimen@
    \if#31
      \overline{\rel@kern{-0.6}\kern\dimen@\macc@nucleus\rel@kern{0.4}\kern\dimen@}%
      \advance\dimen@0.4\dimexpr\macc@kerna
      \let\final@kern#2%
      \ifdim\dimen@<\z@ \let\final@kern1\fi
      \if\final@kern1 \kern-\dimen@\fi
    \else
      \overline{\rel@kern{-0.6}\kern\dimen@#1}%
    \fi
  }%
  \macc@depth\@ne
  \let\math@bgroup\@empty \let\math@egroup\macc@set@skewchar
  \mathsurround\z@ \frozen@everymath{\mathgroup\macc@group\relax}%
  \macc@set@skewchar\relax
  \let\mathaccentV\macc@nested@a
  \if#31
    \macc@nested@a\relax111{#1}%
  \else
    \def\gobble@till@marker##1\endmarker{}%
    \futurelet\first@char\gobble@till@marker#1\endmarker
    \ifcat\noexpand\first@char A\else
      \def\first@char{}%
    \fi
    \macc@nested@a\relax111{\first@char}%
  \fi
  \endgroup
}
\makeatother

\begin{document}
\begin{frontmatter}

\title{A Variational Computational-based Framework for Unsteady Incompressible Flows}


\author[1]{H. Sababha\corref{cor1}}
\ead{haa385@nyu.edu} 
\cortext[cor1]{Corresponding author}

\author[2]{A. Elmaradny}
\author[2]{H. Taha}
\author[1]{M. Daqaq}

\address[1]{New York University Abu Dhabi, Division of Engineering, Abu Dhabi, UAE}
\address[2]{University of California Irvine, Samueli School of Engineering, CA, USA}

\begin{abstract}
Advancements in computational fluid mechanics have largely relied on Newtonian frameworks, particularly through the direct simulation of Navier-Stokes equations. In this work, we propose an alternative computational framework that employs variational methods, specifically by leveraging the principle of minimum pressure gradient, which turns the fluid mechanics problem into a minimization problem whose solution can be used to predict the flow field in unsteady incompressible viscous flows. 

This method exhibits two particulary intriguing properties. First, it circumvents the chronic issues of pressure-velocity coupling in incompressible flows, which often dominates the computational cost in computational fluid dynamics (CFD). Second, this method eliminates the reliance on unphysical assumptions at the outflow boundary, addressing another longstanding challenge in CFD.

We apply this framework to three benchmark examples across a range of Reynolds numbers: $i)$ unsteady flow field in a lid-driven cavity, $ii)$ Poiseuille flow, and $iii)$ flow past a circular cylinder. The minimization framework is carried out using a physics-informed neural network (PINN), which integrates the underlying physical principles directly into the training of the model. The results from the proposed method are validated against high-fidelity CFD simulations, showing an excellent agreement. Comparison of the proposed variational method to the conventional method, wherein PINNs is directly applied to solve Navier-Stokes Equations, reveals that the proposed method outperforms conventional PINNs in terms of both convergence rate and time, demonstrating its potential for solving complex fluid mechanics problems.
\end{abstract}


\end{frontmatter}

\section{\label{sec:Intro}Introduction}
Traditional computational methods in fluid mechanics are predominantly grounded in the Newtonian-mechanics framework, primarily relying on the solution of Navier-Stokes equation (NSE) or one of its various derivatives \citep{gresho1987pressure}. While variational methods have proven effective in other domains (e.g., solid mechanics, quantum mechanics, general relativity) \citep{lanczos2012variational}, the use of variational principles in fluid mechanics has been somewhat limited.  Penfield \cite{Variational_Principles_Fluids_HamiltoN_poF} discussed ``why Hamilton's principle is not more widely used in the field of fluid mechanics". He wrote ``Many important aspects of fluid flow, such as heat flow, turbulence, viscosity, and other irreversible phenomena, cannot be treated". This limitation is due to the fact that the standard form of Hamiltonian's principle of least action cannot directly incorporate non-conservative forces \citep{bretherton1970note, salmon1988hamiltonian, morrison1998hamiltonian,Morrison2020lagrangian}. As a result, variational principles in fluid mechanics have faced challenges in effectively addressing the complexities of real-world scenarios. Although there were several attempts to account for dissipative (e.g., viscous) forces in the principle of least action \cite{Variational_Principles_Fluids_Stochastic,Variational_Principles_NS,Variational_Principles_Fluids_Stochastic_Gomes,Variational_Principles_Fluids_Stochastic2,Variational_Principles_NS_Nonholonomic,Variational_Principles_NS_2n,Variational_Principles_NS_Nonholonomic2,DeVoria_Hamiltonian_JFM}, it remains undetermined how these recent formulations can lead to new computational techniques in fluid mechanics that are different from direct simulation of NSE. 

In a departure from traditional variational principles that are based on least action, Taha et al. \citep{gonzalez2022variational, taha2023minimization} developed a new minimization principle for the incompressible NSE based on Gauss' principle of least constraint \cite{Gauss_Least_Constraint,Papastavridis,Udwadia_Kalaba_Original,udwadia2002foundations,Udwadia_Kalaba_Book,udwadia2023general}: the Principle of Minimum Pressure Gradient (PMPG). This principle asserts that an incompressible flow evolves from one time instant to another such that the total magnitude of the pressure gradient over the domain is minimized. Taha et al. \cite{taha2023minimization} proved that the necessary condition for minimizing the pressure gradient cost is guaranteed to satisfy NSE, which turns a fluid mechanics problem into a pure minimization one. In this approach, the focus is no longer on solving the NSE, but rather on minimizing the cost; and the resulting minimizing solution is guaranteed to naturally satisfy  NSE.

In \cite{taha2023minimization}, Taha et al. employed the PMPG to find analytical solutions to some classical problems in fluid mechanics (e.g. the unsteady laminar flow in a channel and the Stokes’ second problem) without resorting to the solution of the NSE, but rather by formulating each problem as a minimization one. This alternative approach proved to be more insightful in some scenarios. For instance, it was demonstrated that the inviscid version of the PMPG can determine aerodynamic lift over smooth cylindrical shapes, where Euler’s equation struggles to provide unique solutions \cite{gonzalez2022variational}. Additionally, it was illustrated that the flow around a rotating cylinder can be addressed using a straightforward one-dimensional minimization problem, in contrast to Glauert's more cumbersome approach of solving the nonlinear partial differential equations of Prandtl’s boundary layer \cite{shorbagy2024magnus}.  

Most of the existing efforts that employ the PMPG were concerned with theoretical modeling or obtaining analytical solutions. Few studies have implemented the PMPG concept in a computational framework, though focused on steady problems, such as the lid-driven cavity at low Reynolds numbers \cite{alhussein2024principle, alhussein2024variational} and Euler flow past a circular cylinder \cite{atallah2024novel}.

In this work, we aim to exploit the true potential of the PMPG in presenting a general computational framework for unsteady incompressible flows. Following the philosophy of Gauss' principle, the mechanics problem is converted into a minimization problem at every instant of time to determine the best evolution from the current state to the next state. In the unsteady PMPG formulation, we determine the best local acceleration (i.e., the best evolution of the velocity field from the current state) to minimize the pressure gradient cost. Additionally, we present an approach to tackle this infinite-dimensional optimization problem using the framework of physics-informed neural networks (PINNs) \cite{PINNs_Original,PINNs_Karniadakis1,PINNs_Karniadakis2,PINNs_Others2,PINNs_Others3,PINNs_Others4, toscano2024pinns}.

The proposed method offers significant advantages that address several challenges in computational fluid dynamics (CFD). 
\begin{itemize}
    \item First, removing the explicit dependence on pressure eliminates the need for pressure-velocity coupling. The need to solve Poisson equation at every time step is alleviated, which often dominates the computational cost in CFD \cite{jang1986comparison}. 
    \item  Second, the method addresses the inherent issues with outflow boundary conditions. For instance, in most fluid mechanics problems, boundary conditions at the inlet and along the sides of the computational domain are generally known. However, appropriate open boundary conditions at the outflow remains a persistent challenge. This issue has been a subject of ongoing debate since the 1970s \cite{papanastasiou1992new, sani1994resume}. The advent of immense computational power has enabled the extension of computational domains to sufficiently large distances, reducing the impact of outflow boundary distortions. While this approach offers a practical solution for steady-state flows, it is inherently limited for unsteady and periodic flows \cite{nair2024outflow}. In such cases, the periodic nature of the flow extends to infinity, making it impossible to fully eliminate boundary effects within a finite domain. The PMPG framework, by contrast, inherently accommodates such boundary conditions without requiring artificial constraints on the pressure at the outlet, making it particularly advantageous for unsteady and periodic flows.
\end{itemize}

The paper is structured as follows. Section 2 provides a mathematical formulation of the PMPG, shedding light onto its theoretical foundation via Gauss' principle. Section 3 presents the proposed computational framework as a minimization problem. Section 4 demonstrates the proposed approach on benchmark fluid mechanics problems. Section 5 examines the temporal evolution of the flow field. Finally, section 6 presents the conclusions.

\section{The Principle of Minimum Pressure Gradient}\label{sec:FP}
To effectively present the PMPG, a foundational understanding of Gauss' principle of least constraint is essential. This principle, which forms the core of the derivation, is extensively detailed in references \cite{Papastavridis,Udwadia_Kalaba_Original,udwadia2002foundations,Udwadia_Kalaba_Book,udwadia2023general}. Consider the dynamics of $N$  constrained particles, each of a fixed mass $m_i$, whose motion can be described by the generalized coordinates $\mathbf{q}$.  The dynamics of such particles is dictated by Newton's second law as: 
\begin{equation}\label{eq:Newton}
    m_i\; \mathbf{a}_i(\Ddot{q}, \dot{q}, \mathbf{q}) = \mathbf{F}_i+\mathbf{R}_i, \qquad i=1,2, \ldots N,
\end{equation}
where $\mathbf{a}_i$ is the inertial acceleration of the $i$th particle, $\mathbf{F}_i$ are the \textit{impressed} forces acting on the particle, and $\mathbf{R}_i$ are the \textit{constraint} (reaction) forces whose sole role is to preserve the constraints. 

Gauss' principle asserts that the quantity 
\begin{equation}
    Z = \frac{1}{2} \sum_{i=1}^{N} m_i \left( \mathbf{a}_i - \frac{\mathbf{F}_i}{m_i}\right)^2
\end{equation}
is a minimum with respect to the generalized accelerations, $\Ddot{\mathbf{q}}$, at every instant of time.  Equivalently, the cost $Z$ can be written in terms of the constraint forces as:  
\begin{equation}
    Z = \frac{1}{2} \sum_{i=1}^{N}  \left(\frac{\mathbf{R}_i}{m_i}\right)^2,
\end{equation}
implying that the sum of the squares of the constraint forces must be a minimum. 

In the absence of constraint forces, $\mathbf{R}_i = 0$, the \textit{optimal} motion that minimizes $Z$ is easy to find: $\mathbf{a}_i = \frac{\mathbf{F}_i}{m_i}$, implying that a particle follows the impressed force applied to it. This is called the \textit{free} motion. However, in a constrained setting, Nature acts like a mathematician (in the words of Gauss' \cite{Gauss_Least_Constraint}) who picks, at each instant of time, an acceleration for the constrained system that minimizes, in a weighted least-squares sense, the deviation between the acceleration of the free motion and the constrained system. This means that, the particle adjusts its path only to the extent necessary to meet the constraints, ensuring the least possible deviation from the unconstrained trajectory (i.e., its free/natural motion).

Gauss' principle, therefore, turns the mechanics problem governed by Newton's Equation (\ref{eq:Newton}) into the instantaneous minimization problem: $\underset{\mathbf{a}_i}{\min} \;\; Z(\mathbf{a}_i)$ subject to the imposed constraints (written in terms of accelerations $\mathbf{a}_i$).

Taha et al. \cite{taha2023minimization} extended Gauss' principle of least constraint to a continuum of fluid particles in an incompressible fluid, whose motion is typically described by the incompressible NSE:
\begin{align}
    \frac{\partial \mathbf{u}}{\partial t} + \mathbf{u} \cdot \nabla\mathbf{u} &= -\nabla p + \nu \nabla^2\mathbf{u}, \label{NSEa} \\
    \nabla \cdot \mathbf{u} &= 0, \label{NSEb}
\end{align}
where $\mathbf{u}$, is the fluid velocity vector, $p$ is the pressure, and $\nu$ is the kinematic viscosity. The problem is closed with the boundary conditions (BC)
\begin{align}
    \mathbf{u} &= \mathbf{w} \quad   \text{for} \; t>0  \quad \text{on} \quad \Gamma,
\end{align}
for a given velocity $\mathbf{w}$ defined on the boundary $\Gamma$, and the initial conditions (IC)
\begin{align}
    \mathbf{u(x, 0)} = \mathbf{u}_0(\mathbf{x}),
\end{align}
for some  initial velocity field satisfying $\nabla . \mathbf{u}_0 = 0$.

Equation (\ref{NSEa}) represents a balance of forces applied on the moving fluid in a domain $\Omega \in R^3$. The left-hand side of the Equation is the total acceleration of the fluid, incorporating both local and convective components, while the right-hand side comprises the forces acting on the fluid, which can be decomposed into either impressed or constraint forces. Equation (\ref{NSEb}) represents a mass balance often referred to as continuity.  

Several efforts \citep{gresho1987pressure,Moin_Incompressible1,Morrison2020lagrangian,gonzalez2022variational,DeVoria_Hamiltonian_JFM}  demonstrated that, in incompressible flows, pressure is a Lagrange multiplier that enforces the continuity constraint. In other words, the pressure gradient field acts as a constraint force, primarily enforcing continuity. Hence, by applying Gauss' principle of least constraint to the dynamics of incompressible fluids, and classifying the pressure gradient as a constraint force, we can write the Gaussian cost in Eulerian coordinates as:
\begin{equation}
    \mathcal{A} = \displaystyle \frac{1}{2} \rho  \int_{\Omega} \left(\frac{\partial \mathbf{u}}{\partial t} + \mathbf{u} \cdot \nabla \mathbf{u} - \nu \Delta \mathbf{u} \right)^2 d\mathbf{x}, \label{actionS}
\end{equation}
subject to the continuity constraint
\begin{equation}
    \nabla\; . \; \mathbf{u} = 0,  
    \label{actionConstraint}
\end{equation}
and any boundary conditions defined on $\Gamma$. Note that the cost $\mathcal{A}$ is nothing but the $L^2$ norm of the pressure gradient field: $\mathcal{A} = \frac{1}{2} \int_\Omega |\mathbf\nabla p|^2 d\mathbf{x}$.  Taha et al. \cite{taha2023minimization} proved that if the acceleration $\frac{\partial \mathbf{u}}{\partial t}(\mathbf{x})\equiv \mathbf{u}_t$ is differentiable in $\Omega\subset\mathbb{R}^3$ and minimizes the cost functional $\mathcal{A}(\mathbf{u}_t)$, given in Eq. (\ref{actionS}), subject to the constraint 
\begin{equation}
    \nabla\; . \; \mathbf{u}_t = 0,  
    \label{actionConstraint_acceleration}
\end{equation}
and the normal flow condition 
\[ \mathbf{u}(\mathbf{x},t) \cdot \mathbf{n}=\mathbf{g}(\mathbf{x},t) \;\;  \text{for all} \; \mathbf{x}\in\Gamma, \;\; t\in\mathbb{R},\]
where $\mathbf{n}$ is the normal to the boundary $\Gamma$, for some $\mathbf{g}$ differentiable in $t$, then $\mathbf{u}_t(\mathbf{x})$ must satisfy the NSE (\ref{NSEa}) \cite{taha2023minimization}.

 This implies that the NSE is the first-order necessary condition for minimizing the cost in Equation (\ref{actionS}) with respect to the acceleration $\mathbf{u}_t$. Therefore, following the philosophy of Gauss’ principle, the PMPG turns the incompressible fluid dynamics problem, governed by the NSE (\ref{NSEa}), into an instantaneous minimization problem in terms of the acceleration field $\mathbf{u}_t$. As such, the local acceleration is determined by minimizing the cost (\ref{actionS}) with respect to $\mathbf{u}_t$ subject to the continuity constraint, written in terms of the acceleration: Equation (\ref{actionConstraint_acceleration}). 

\section{Computational Optimization Framework for Fluid Mechanics Simulations}
In this section, we present a computational framework for the PMPG that enables tackling complex problems where analytical solutions cannot be obtained. To achieve this goal, we employ the concept of  PINNs, which formulates the physics problem into an unconstrained minimization framework. The subsequent sections provide detailed explanations of the proposed approach.

We begin by presenting the unsteady formulation based on the principle of minimum pressure gradient, denoted by PMPG-PINN.  The PMPG-PINN is fundamentally different from conventional PINNs as applied to NSE in several aspects\footnote{Throughout this manuscript, we loosely utilize the term \textit{conventional PINNs} to refer to the application of PINNs to directly solve NSE as presented in Refs. \cite{PINNs_Karniadakis1, PINNs_Karniadakis2, PINNs_Original}}. First, in contrast to conventional PINNs whose cost function is a weighted sum of the residuals of Equations (\ref{NSEa}, \ref{NSEb}), we use the PMPG cost function (\ref{actionS}) subject to the continuity constraint (\ref{actionConstraint_acceleration}). Second, unlike conventional PINNs, the proposed PMPG-PINN formulation does not require constructing a separate network for the pressure, since the cost, Equation (\ref{actionS}), does not include pressure---an advantage that provides savings in training computational time. Finally, the proposed approach solves a minimization problem at every instant of time to obtain the optimal instantaneous acceleration $\mathbf{u}_t(\mathbf{x})$ as a function of space. In contrast, conventional PINNs finds $\mathbf{u}(\mathbf{x},t)$ and $p(\mathbf{x},t)$ that minimize the sum of the residuals from all instants. In particular, throughout the proposed approach, the neural network (NN) is only used as a function approximator to turn the infinite-dimensional optimization (calculus of variations) problem into a finite-dimensional ones (over the parameters of the NN). 

\subsection{Unsteady PMPG-PINN Formulation}
We rely on neural networks (NNs) as function approximators to represent the spatial variation of the unknown acceleration $\mathbf{u_t}(\mathbf{x})$ in terms of finite number of parameters $\theta$, thereby converting the infinite-dimensional optimization problem of the PMPG over $\mathbf{u_t}(\mathbf{x})$ into a finite-dimensional minimization problem over $\theta$.  

\begin{figure}
\centering
\includegraphics[width=1.15\textwidth]{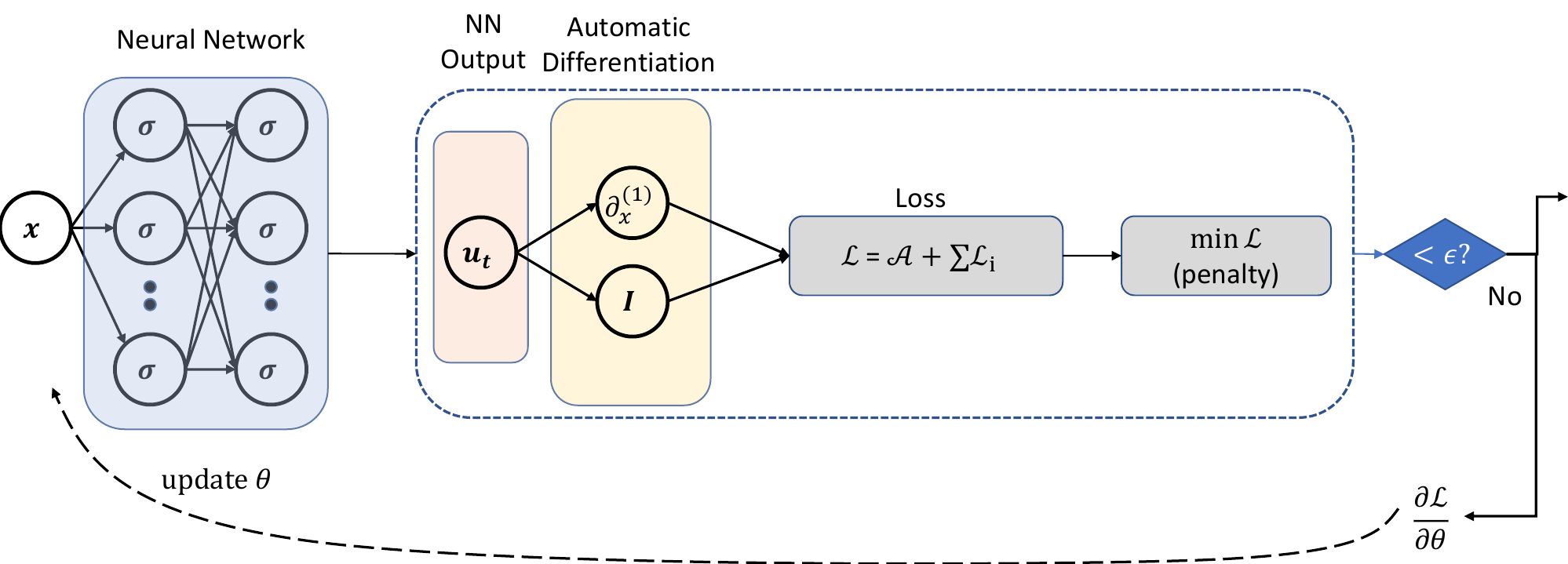}  
\caption{Schematic of a physics-informed neural network based on the unsteady PMPG formulation}\label{PINNSchematicpmpg}.
\end{figure}

As shown in Fig. \ref{PINNSchematicpmpg}, the NN is a fully-connected feed-forward network composed of $L$ multiple-hidden layers. It takes a concatenation of the state-space $z^0 = \mathbf{x}$ as an input, and outputs a guess for the unknown variables, $\mathbf{u_t}$.  Each layer creates data for the next layer through a tensorial nested transformation of the form \citep{svozil1997introduction}:
\begin{equation}
 z^l = \sigma^l (W^l.z^{l-1} + b^l), \quad l = 1, 2, \ldots, L, 
\end{equation}
where the functions $\sigma^l$ are called activation functions. These can be chosen based on the nature of the problem.  The variables $W^l$ and $b^l$ denote, respectively, the weights and biases of each NN layer, $l$. These are the parameters of the NN: $\theta=[W^l,b^l]$, which completely determine an approximation of the sought function $\mathbf{u_t}$. The parameters $W^l$ and $b^l$ are updated after every iteration (epoch) by minimizing a loss function, $\mathcal{L}$, with respect to those variables. In conventional PINNs, the loss function $\mathcal{L}(\theta)$ measures the difference between the predicted solution and the target solution of NSE. When the change in the residual between successive iterations is less than a predefined threshold, say $\epsilon$, the training stops and the output of the last iteration is considered the minimum of the value $\mathcal{L}$. 

When considering $\mathbf{u}_t$ as the primary output variable from the NN, we need to express the continuity constraint and boundary conditions in terms of $\mathbf{u_t}$. One can easily show that, if the initial flow field $\mathbf{u}_0$ satisfies the divergence-free condition (i.e. $\nabla . \mathbf{u}_0 = 0)$, then the condition $\nabla \cdot \mathbf{u}_t = 0$ ensures that $\nabla \cdot \mathbf{u} = 0$ at the subsequent time step. Moreover, by imposing zero boundary conditions on $\mathbf{u}_t$, the boundary conditions on the velocity field are preserved at each subsequent time step as well. Consequently, the optimization problem is to minimize the pressure gradient cost subject to two constraints: the divergence-free condition $\nabla \cdot \mathbf{u}_t = 0$ and the BC: $\mathbf{u}_t = 0$ on $\Gamma$.

These constraints are considered in the PMPG-PINN formulation using the penalty method. That is, the loss function is written as:
\begin{equation}
    \mathcal{L}(\theta)  = \mathcal{A} + \mu_c  \mathcal{L}_c + \mu_b  \mathcal{L}_{bc},
    \label{loss_PMPG}
\end{equation}
where the loss $\mathcal{L}_{c}$, weighted by $\mu_c$, penalizes violation of the continuity constraint, and is defined as: 
\begin{equation}\label{eq:Continuity_Loss}
    \mathcal{L}_{c} = \frac{1}{N_C} \sum_{i}^{N_C}  g_i(\mathbf{x})^2, \quad g = \nabla\,. \mathbf{u}_t,
\end{equation}
where $N_C$ is the number of points in the domain.

On the other hand, the loss $\mathcal{L}_{bc}$, weighted by $\mu_b$, penalizes violation of the boundary conditions: 
\begin{equation}\label{eq:lossBC}
    \mathcal{L}_{bc} = \frac{1}{N_B} \sum_{i}^{N_B}  (\mathbf{u}_t(\mathbf{x}))^2,
\end{equation}
where $N_B$ is the number of collocation points over the boundary, $\Gamma$, of the domain. In some cases, boundary conditions can be imposed using an anstaz function on the output of the neural network, thus removing the penalty term $\mathcal{L}_{bc}$ from the loss function. 

\begin{figure}
\centering
\includegraphics[width=0.75\textwidth]{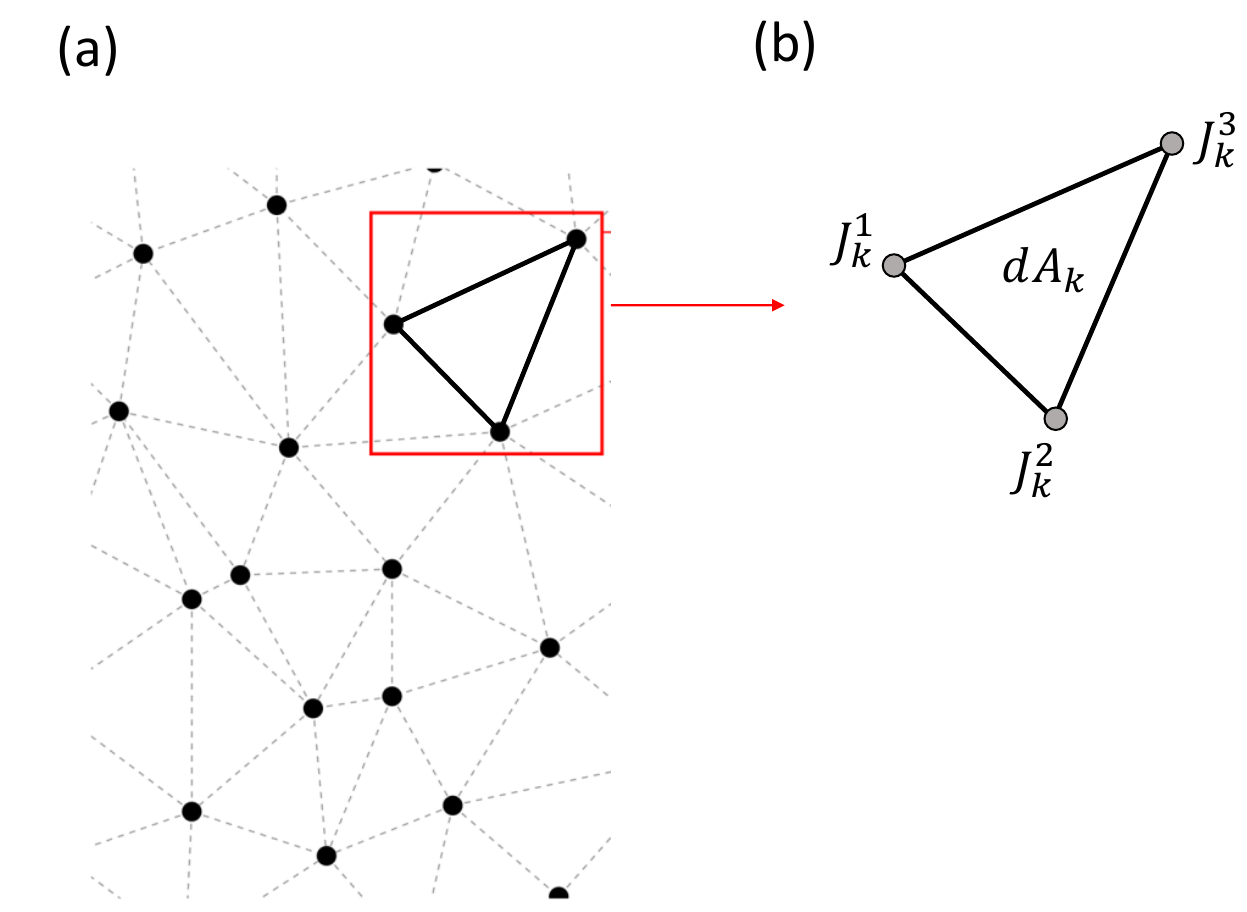}  
\caption{Illustration of the numerical integration algorithm towards 2D randomly distributed sample points.}\label{delIntg}.
\end{figure}

Finally, the integral $\mathcal{A}$ can be computed based on the distribution of points within the domain. For uniform sampling, the integral can be written as a sum using the mean rule, leading to
\begin{equation}
    \mathcal{A} = \frac{A}{N_c} \sum_{i}^{N_c}     J_i,
\end{equation}
Here, $A$, is the area of the domain, and, $J$, is the discretized operator used to approximate the integrand in Equation \eqref{actionS}: 
\begin{equation}
    J = \displaystyle \frac{1}{2}\rho \left(\mathbf{u}_t + \mathbf{u} \cdot \nabla \mathbf{u} - \nu \Delta \mathbf{u} \right)^2.
\end{equation}
In the case of arbitrarily distributed sample points, the computational domain is firstly discretized into triangles using the Delaunay Triangulation algorithm \cite{lee1980two}, as shown in Fig. \ref{delIntg}. Consequently, the cost integral $\mathcal{A}$ is computed using 
\begin{equation}
    \mathcal{A} = \sum_{i}^{N_t}     \bar{J}_i\,\text{d}A_i ,
\end{equation}
where $N_t$ is the number of formed triangles in the domain, $\text{d}A_i$ is the area of each triangle and $\bar{J}_i$ is the mean value of  $J$ at the vertices of the respective triangle.  

In the PMPG-PINN framework, the minimization of the loss function can be achieved when the constraint terms $\mathcal{L}_i$ are zero, while the cost function $\mathcal{A}$ remains non-zero. As a result, an optimization scheme that simply minimizes the sum of the objective and constraint losses often converges to a suboptimal solution rather than a true minimum \citep{nocedal2006penalty}. To overcome this limitation, we use the penalty method \citep{nocedal2006penalty}, which replaces constant penalty terms with a sequence of outer iterations, progressively increasing the penalty applied to the constraints. In this approach, the updated loss function at the $k^{\rm{th}}$ iteration is expressed as:
\begin{equation}
    \mathcal{L}(\theta)  = \mathcal{A} + \sum_i \mu_i^{k}  \mathcal{L}_{i},
    \label{loss_PMPGPenalty}
\end{equation}
where $\mu_i^k = \gamma \,\mu_i^{k-1} $, and $\gamma$ is a constant value chosen depending on the problem being solved.    

\subsection{Conventional PINN Formulation}
Since a direct comparison with conventional PINNs may not be conclusive, due to the fundamental difference between the two formulations, we introduce a modified version of PINNs where the residuals of the NSE (\eqref{NSEa},\eqref{NSEb}) are minimized at every instant of time. Here, a neural network model is required for all unknowns in Equations \eqref{NSEa} and \eqref{NSEb} at the given instant: $\mathbf{u}_t(\mathbf{x})$, $p(\mathbf{x})$.  

\begin{figure}
\centering
\includegraphics[width=1.15\textwidth]{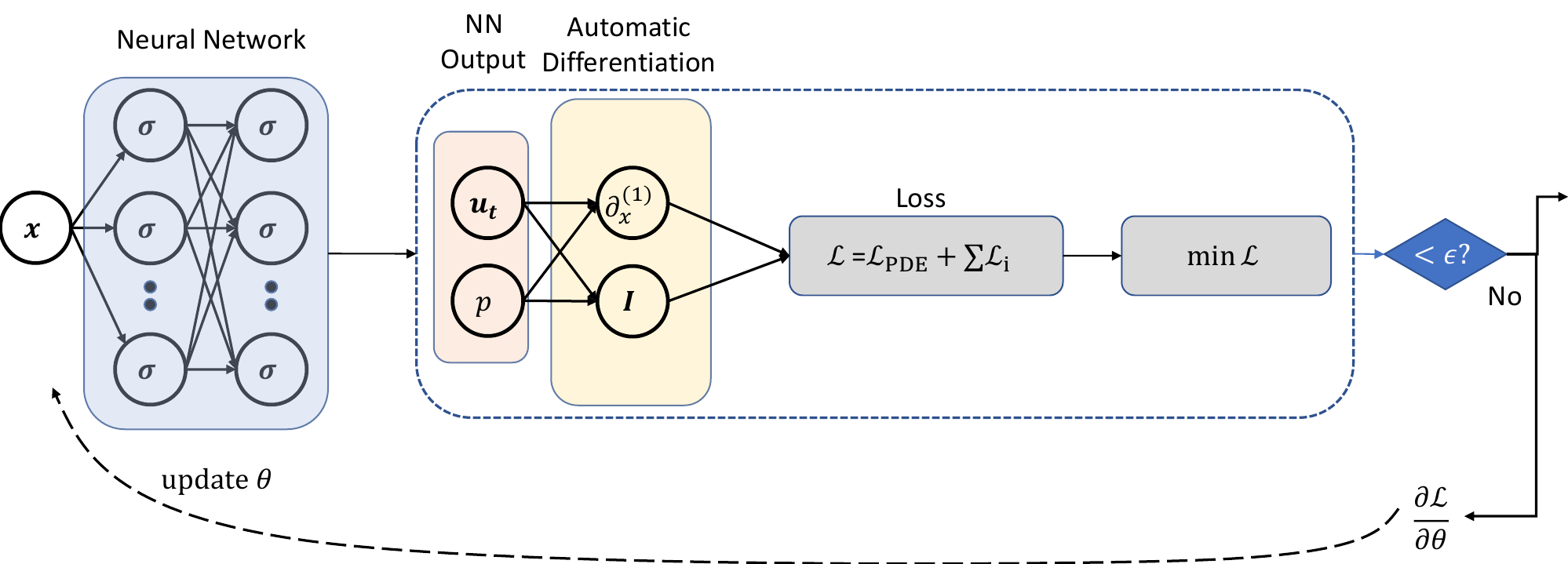}  
\caption{Schematic of a physics-informed neural network based on unsteady NS formulation}\label{PINNSchematicNS}.
\end{figure}

Figure \ref{PINNSchematicNS} shows the structure of the conventional PINN formulation. The NN takes a concatenation of the state-space $z^0 = \mathbf{x}$ as an input, and outputs a guess for the unknown variables ($\mathbf{u_t},p$). 
The loss is then formed by three terms as:
\begin{equation}
    \mathcal{L}(\theta)  =\mu_p \mathcal{L}_{PDE} + \mu_c  \mathcal{L}_c + \mu_b  \mathcal{L}_{bc},
    \label{loss_PMPG2}
\end{equation}
 where the first term $\mathcal{L}_{PDE}$, weighted by $\mu_p$, represents the residual in the momentum Equation \eqref{NSEa}, and is given by:
 \begin{equation}
    \mathcal{L}_{PDE} = \frac{1}{N_C} \sum_{i}^{N_C} f_i(t, \mathbf{x})^2, \quad f =   \frac{\partial \mathbf{u}}{\partial t} + (\mathbf{u} \cdot \nabla)\mathbf{u} 
   +\nabla p - \nu \nabla^2\mathbf{u}.
\end{equation}
 The remaining loss terms, $\mathcal{L}_c$ and $\mathcal{L}_{bc}$, represent constraints on the divergence and boundary conditions, respectively, defined in Equations (\ref{eq:Continuity_Loss},\ref{eq:lossBC}). 
 
This setup ensures that both approaches---the PMPG-PINN and the conventional PINN---are comparable under similar conditions, enabling a fair assessment of their performance.

\section{Numerical Examples}
To demonstrate the effectiveness of the proposed approach, we consider three benchmark examples in fluid mechanics: $i)$ the Poiseuille flow (the unsteady laminar flow in a channel of uniform cross-section) \cite{thomas1953stability}, $ii)$ the unsteady flow inside a lid-driven cavity \cite{koseff1984lid}, and $iii)$ the separating flow past a circular cylinder \cite{fornberg1980numerical}. To reduce complexity, and without loss of generality, we focus solely on two-dimensional flows.

\begin{figure}[!tb]
\centering
\includegraphics[width=0.75\textwidth]{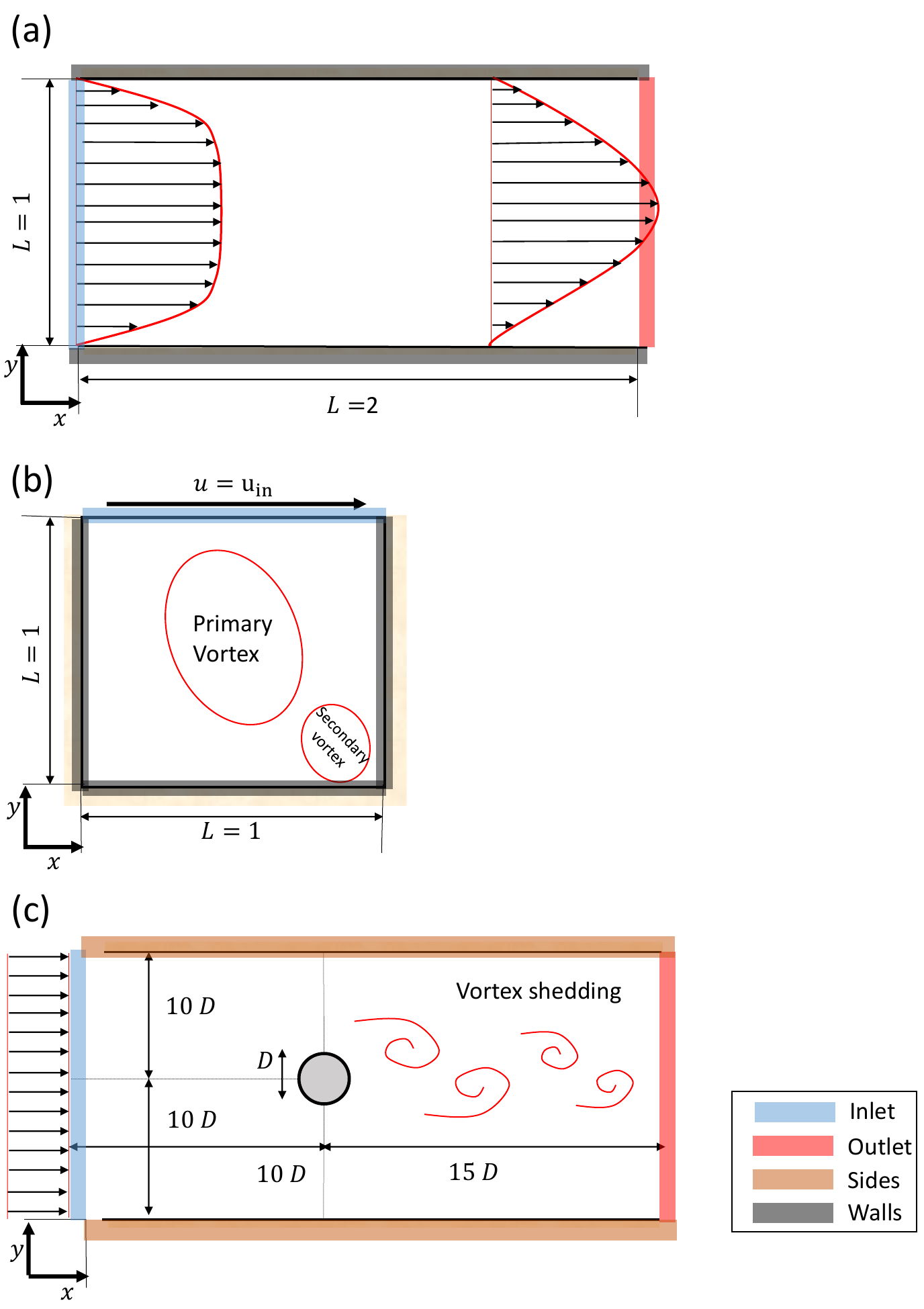}
\caption{Schematic diagram of the two-dimensional (a) Poiseuille flow, (b) lid-driven cavity problem, and (c) the flow past a circular cylinder.}\label{schematic}.
\end{figure}

The solution to all problems follow the same procedure which starts by generating an initial (reference) flow field using a conventional CFD solver (see Appendix for details). It is important to highlight that the initial flow field does not necessarily need to be generated from high-fidelity simulations. The only requirement is that the flow is divergence-free and satisfies the boundary conditions; it must be kinematically admissible, but not necessarily dynamically correct. This reference flow is fed as an initial condition into the PMPG-PINN minimization framework, which computes the acceleration, $\mathbf{u}_t$, that minimizes the pressure gradient cost, $\mathcal{A}$. For validation purposes, the predicted $\mathbf{u}_t$ is compared to reference values obtained using a high-fidelity CFD solver.

\subsection{Two-dimensional Poiseuille Flow} \label{poiseuille}
The first benchmark example serves to $i)$ detail the computational procedure, $ii)$ validate its output against high-fidelity CFD simulations, and $iii)$ compare its performance with conventional PINNs. As shown in Fig.~\ref{schematic}(a), we consider predicting the spatial evolution at a given random time instant of a laminar flow in a channel defined over the domain $[0, 1] \times [0, 2]$. To implement the PINN algorithm, a reference flow is first generated using CFD ensuring a divergence free flow and satisfying the boundary conditions. In the CFD, the boundary conditions are $i)$ no-slip conditions on the walls, and $ii)$ an inlet velocity profile $U_0(y)$ at the inlet, and $iii)$ a zero pressure at the outlet. Recall that, since PMPG-PINN is pressure independent, the boundary condition on the outlet pressure is no longer needed for the convergence of the solution procedure. Here, to avoid discontinuities at the corners, we replace the commonly used uniform inlet flow, with a nonuniform profile that takes the shape of the following function:
\begin{equation}
    U_0(y) = 1 - \frac{\text{cosh}\left(C_0(y-0.5)\right)}{\text{cosh}(0.5 C_0))}
\end{equation}   
where $C_0 = 15$. Based on the maximum inlet velocity and the kinematic viscosity, the Reynolds number is equal to 100. 

The PINN takes a uniform sampling points $\mathbf{x} = (x, y)$ as inputs, and outputs the local acceleration $\mathbf{u_t} = (u_t, v_t)$ based on the reference flow. To reduce skewness of input data, and facilitate the neural network's learning process, the input domain is scaled to a range between 0 and 1 \cite{chollet2021deep}. Then, to force the output solution to satisfy the boundary conditions, we construct a NN for ($\hat{u_t}, \hat{v_t}$) instead of $(u_t, v_t)$ and determine the former as:  
 \begin{align}
\hat{u}_t &=  x_s\,y_s\,(1-y_s)u_t,\\
\hat{v}_t &=  x_s\,y_s\,(1-y_s)v_t,
 \end{align}
where $x_s \text{ and } y_s$ are the scaled values of the inputs $x \text{ and } y$.
 
During network training, the temporal continuity constraint is penalized with increasing weights, namely, $\mu_i^{(1)} = 5$, $\gamma = 2$, for  6 outer iterations. That is, the boundary conditions are automatically satisfied, and the continuity constraint is enforced by the penalty method. The network was implemented with $nn = 32$ neurons per layer, $nl = 10$ layers, and the number of collocations points is $31250$, which resulted in acceptable residual errors.

\begin{figure}[!tbh]
\centering
\includegraphics[width=\textwidth]{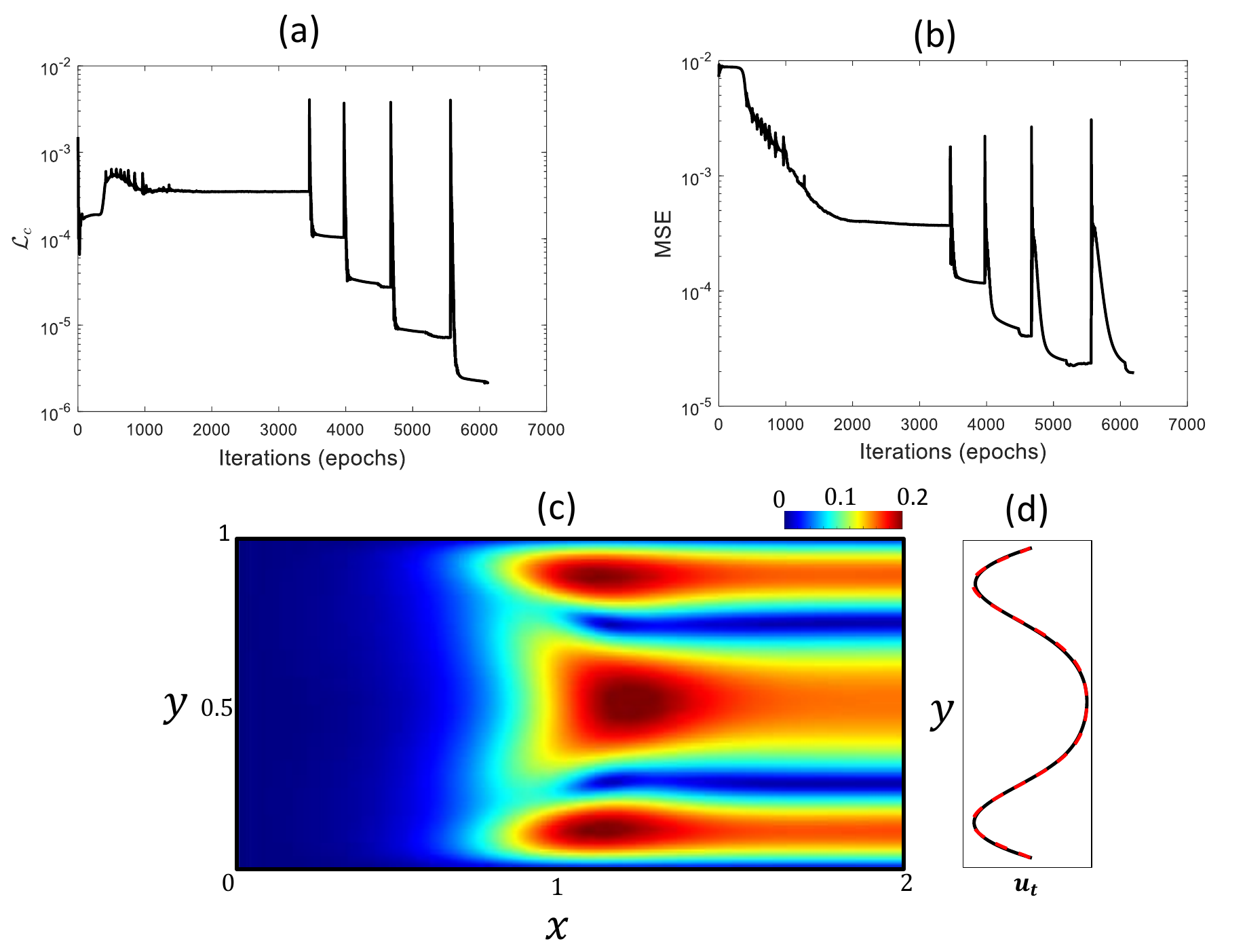}   
\caption{Results from the minimization training. (a) Residual loss function, $\mathcal{L}_c$. (b) Mean squared error of the predicted $\mathbf{u}_t$ and the reference values, (c) color map of the predicted $\mathbf{u_t}$, and (d) comparison of the predicted $\mathbf{u_t}$ (line) with reference value (dashed lines) at the outlet boundary. }\label{fig:mainPois}.
\end{figure}

The results of the minimization process are presented in Fig.~\ref{fig:mainPois}. Panel (a) shows the evolution of the residual loss $\mathcal{L}_c$ with the number of epochs. As the number of outer iterations increases, the penalty associated with the constraint also rises, resulting in a stair-case decrease in the residual loss function, $\mathcal{L}_c$. Ultimately, the residual reach an acceptable value of less than $10^{-6}$. The accuracy of the predicted $\mathbf{u}_t$ is demonstrated in Fig.\ref{fig:mainPois}(b), where the mean square error (MSE) summed over the whole domain is plotted against the number of iterations.  The MSE is based on the difference between the values of $\mathbf{u_t}$ predicted using the PMPG-PINN and those obtained using the high-fidelity CFD simulations. The plot shows that the error converges to a low value, confirming the reliability of the predictions. The $\mathbf{u_t}$ resulting from the PMPG-PINN simulation is depicted in Fig.~\ref{fig:mainPois}(c), where the minimum value of $\mathcal{A}$ resulting from the PMPG-PINN was computed to be $\mathcal{A}=0.049$, closely matching that computed using the CFD model where  $\mathcal{A}$ was found to be $\mathcal{A}=0.05$. 

Finally, Fig.~\ref{fig:mainPois}(d) compares $\mathbf{u_t}$ at the outlet with the high-fidelity results, clearly demonstrating excellent agreement between the two. This observation underscores a significant advantage of the approach. Notably, despite the omission of pressure from the formulation, the problem remained well-posed, and no additional boundary conditions were necessary on the outlet. This advantage is appreciated when we recall that the pressure-velocity coupling is the most computationally expensive step in an incompressible flow solver.

It may be instructive to investigate the performance of the PMPG approach in comparison to the conventional PINN formulation. For a fair comparison, we used one outer iteration and set the weight on the constraints to $\mu = 25$. Both methods were tested using the same architecture as in the previous example, with parameters $nn = 32$, $nl = 10$ and $N_B = 31250$. During the training process, we traced the mean square error between the PMPG-PINN and the reference high-fidelity solution.

\begin{figure}[!tbh]
\centering
\includegraphics[width=0.75\textwidth]{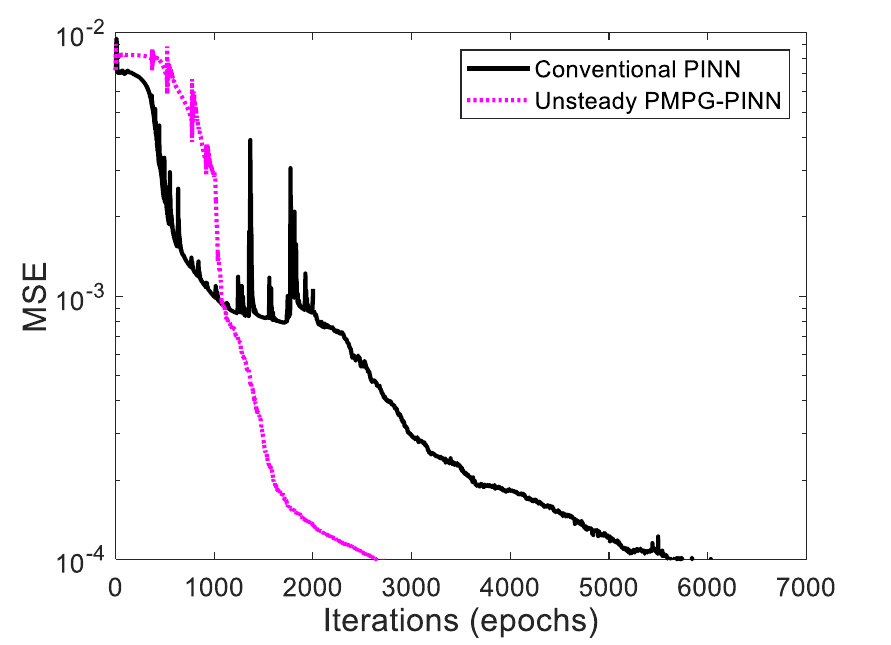}
\caption{Comparison of the mean square error of the PMPG and least-squares formulations. }\label{fig:compare}.
\end{figure}

Figure~\ref{fig:compare} presents the convergence of the mean squared error with iterations for both approaches, which clearly shows that the PMPG-PINN outperforms the conventional PINN in both convergence time and rate. The convergence time, measured in terms of time per epoch, is shorter in the PMPG formulation. Specifically, we observed that the time per epoch is 20\% less than that of the conventional PINN. Additionally, the figure shows that convergence is achieved using fewer epochs. This faster convergence can be attributed to two main factors: First, the removal of pressure from the equations eliminates the need for auto-differentiation, which is typically computationally intensive and time-consuming. Second, removal of pressure also reduces the solution space, which simplifies the search for the optimal minimum, allowing faster convergence.

\subsection{Two dimensional Lid-Driven Cavity Flow}
The lid-driven cavity problem is a fundamental benchmark in CFDs due to its simple geometry, but interesting fluid dynamics. The Reynolds number plays a key role in characterizing the cavity flow. It is generally accepted that the flow remains laminar for $Re < 7000$ \cite{peng2003transition}. However, to the best of the authors' knowledge, there have been no reports of PINNs accurately solving the cavity flow problem at a relatively high Reynolds numbers (e.g., 1000 and higher) \cite{he2023artificial, wang2023solution}. To this end, we examine the capability of the new formulation to predict the flow field across both low and high Reynolds numbers within the laminar regime. 

To achieve this goal, we run the optimization algorithm at three different Reynolds numbers. Here, the maximum axial velocity remains the same in all simulations, and the kinematic viscosity is chosen so that the flow remains laminar with a Reynolds number of $Re =$ 25, 500, 5000. As shown in Fig. \ref{schematic}(b), the computational domain is defined on the area [0, 1] x [0, 1].  A no-slip condition is enforced on the four walls, where three walls are fixed while the upper one is assumed to move to the right with a parabolic velocity profile. 

The PINN takes a uniform sampling points $\mathbf{x} = (x, y)$ as inputs, and outputs the local acceleration $\mathbf{u_t} = (u_t, v_t)$. To force the output solution to satisfy the boundary conditions, we use a similar trick and construct a NN for ($\hat{u_t}, \hat{v_t}$) instead of $(u_t, v_t)$ and determine the former as:  
 \begin{align}
 \hat{u}_t &=  x(1-x)y(1-y)u_t,\\
 \hat{v}_t &=  x(1-x)y(1-y)v_t.
 \end{align}

\begin{figure}[!tb]
\centering
\includegraphics[width=\textwidth]{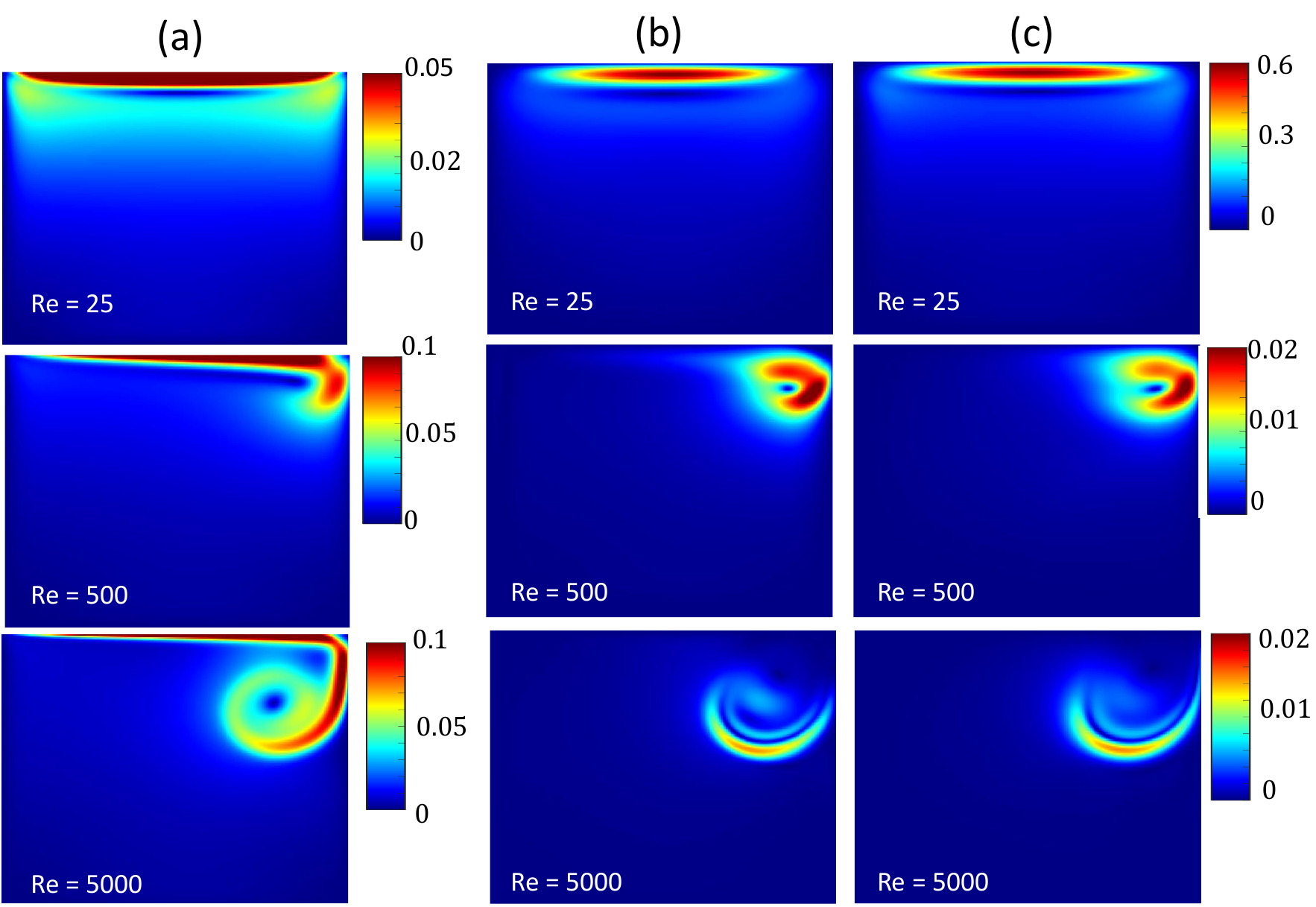}
\caption{Results from Lid-Cavity training at different Reynolds numbers. (a) Initial flow field, $\mathbf{u}$. (b) Local acceleration, $\mathbf{u_t}$, obtained using high-fidelity CFD models. (c) Predicted $\mathbf{u_t}$ using the unsteady PMPG-PINN formulation.}\label{fig:mainLid}.
\end{figure}

The results are presented in Fig.~\ref{fig:mainLid}. Panel (a) shows the initial reference flow field, $\mathbf{u}$, used as an input for the PMPG-PINN formulation. Panel (b) shows the predicted acceleration $\mathbf{u}_t$ using the high-fidelity CFD solver, while panel and (c) shows the predicted acceleration $\mathbf{u}_t$ as obtained using PMPG-PINN. Results demonstrate excellent agreement between the two frameworks across all considered Reynolds numbers: $Re=$ 25, 500, and 5000. 

\subsection{Flow Past A Circular Cylinder}
Our third example considers flow over a bluff body---an important class in engineering due to its complex behavior, characterized by flow separation. A well-studied example is the flow past a circular cylinder, which serves as a benchmark for analyzing vortex dynamics. At low Reynolds numbers, flow separation occurs without vortex shedding, but as the Reynolds number exceeds a critical threshold, periodic vortex shedding develops, forming a von Kármán vortex street \cite{king1977review}. 

This benchmark example serves two primary objectives. First, it investigates the efficacy of the framework for an essentially periodic flow, such as predicting vortex shedding at $Re = 100$. Second, it demonstrates the robustness of the framework, even when the computational domain is truncated to less than $8D$ downstream of the cylinder; a scenario where conventional CFD approaches struggle to produce accurate results.

As shown in Fig. \ref{schematic}(c), the computational domain in the first demonstration spans [-10, 15] x [-10, 10].  For the reference flow obtained using CFD, a no-slip condition is enforced on the cylinder wall, a constant velocity is enforced on the inlet, as well as along the upper and lower boundaries of the computational domain. A zero pressure condition is enforced at the outlet. Since the PMPG-PINN framework is pressure-independent, no pressure boundary condition is specified at the outlet when solving the problem using PMPG-PINN.

\begin{figure}[!tb]
\centering
\includegraphics[width=0.4\textwidth]{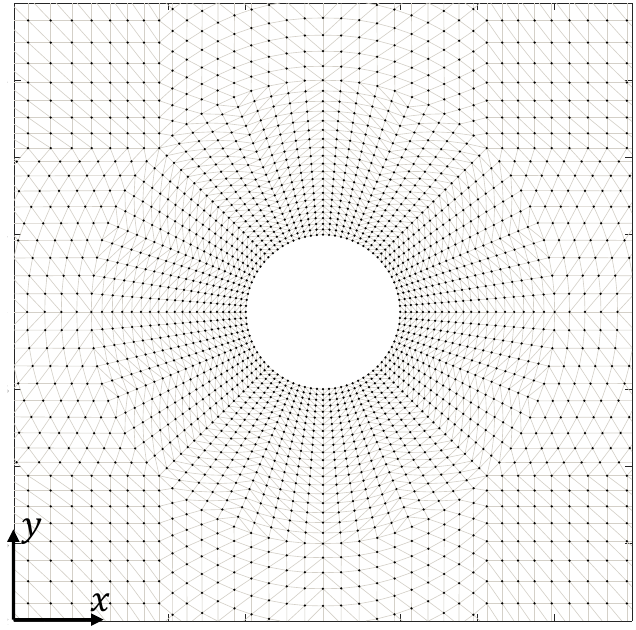}
\caption{Domain of the flow past a circular cylinder. Dots show the control points, $N_C$, and lines show the Delaunay triangulation.}\label{fig:dmCP}.
\end{figure}

In this problem, we use non-uniformly distributed collocation points, as illustrated in Fig. \ref{fig:dmCP}. Near the cylinder wall, where high gradients are expected, the sampling density is increased, while in regions farther from the cylinder, where gradients are smaller, a coarser distribution of points is applied. Similarly, we scale $\mathbf{x}$ to reduce skewness of input data. The NN outputs the local acceleration $\mathbf{u_t} = (u_t, v_t)$. To force the output solution to satisfy the far-field boundary conditions (inlet, upper and lower boundaries), we use our usual trick of constructing a NN for ($\hat{u_t}, \hat{v_t}$) instead of $(u_t, v_t)$ and determine the former as:  
 \begin{align}
 \hat{u}_t &=  x_sy_s(1-y_s)u_t,\\
 \hat{v}_t &=  x_sy_s(1-y_s)v_t,
 \end{align}
Note that the above equations do not enforce the no-slip condition on the cylinder wall. As such, a soft constraint imposing no-slip condition, $\mathcal{L}_{bc}$, according to Equation \eqref{eq:lossBC} is added to the overall loss function, $\mathcal{L}$.

\begin{figure}[!tb]
\centering
\includegraphics[width=1.0\textwidth]{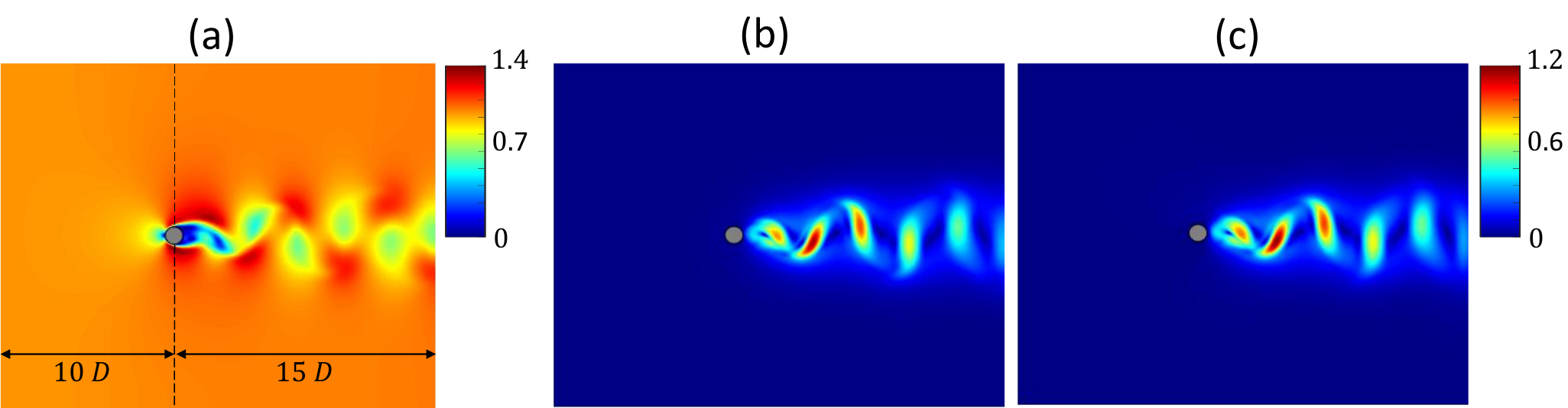}
\caption{Results from the flow past a circular cylinder at $Re = 100$. (a) Reference flow field, $\mathbf{u}$. (b)  Local acceleration, $\mathbf{u_t}$, obtained using high-fidelity CFD models. (c) Local acceleration, $\mathbf{u_t}$, obtained the unsteady PMPG-PINN formulation.}\label{fig:mainFPC}.
\end{figure}

The results are presented in Fig.~\ref{fig:mainFPC}. Panel (a) shows the initial reference flow field, $\mathbf{u}$, used as an input for the PINN optimization. Panels (b) and (c) then compare the  $\mathbf{u}_t$ obtained using CFD to that predicted by PMPG-PINN. The results demonstrate excellent agreement between the PMPG-PINN values and those obtained using the high-fidelity simulations. 

\begin{figure}[!tb]
\centering
\includegraphics[width=1.0\textwidth]{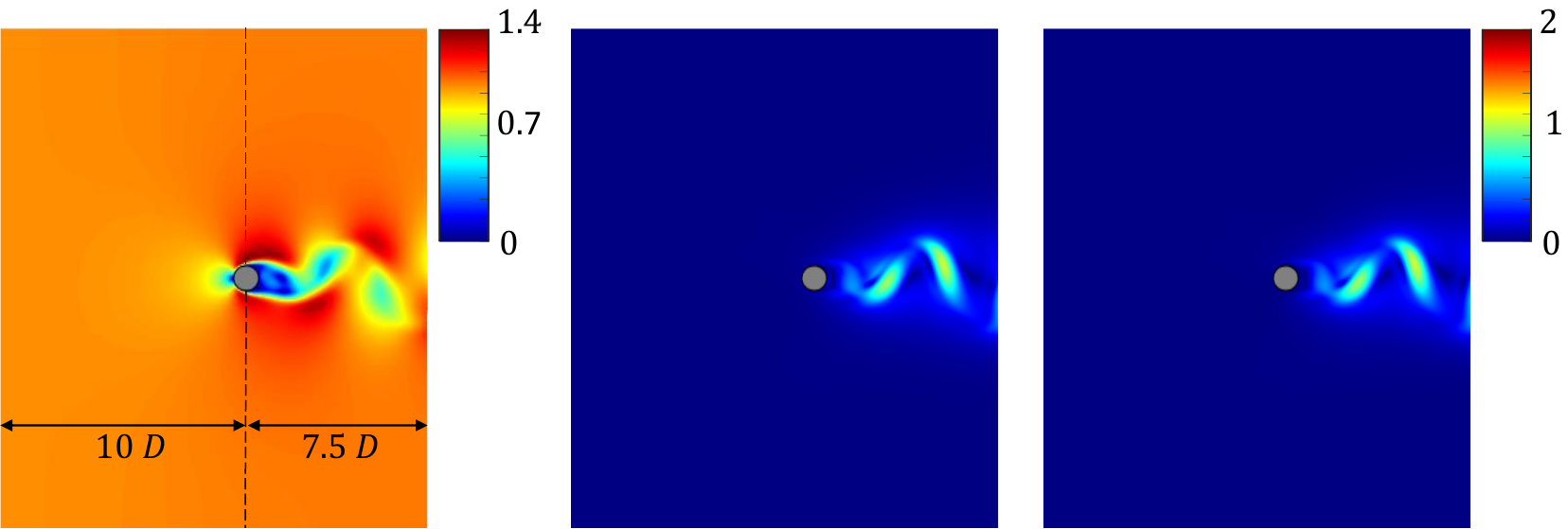}
\caption{Results from the flow past a circular cylinder at $Re = 100$. (a) Reference flow field, $\mathbf{u}$. (b) Local acceleration, $\mathbf{u_t}$, obtained using high-fidelity CFD simulation. (c) Local acceleration, $\mathbf{u_t}$, predicted using the unsteady PMPG formulation.}\label{fig:mainFPC_truncated}.
\end{figure}

In the second demonstration, we consider a truncated computational domain, where the outflow boundary is placed close to the cylinder, within the region where vortices are still forming. The reference flow for this domain is shown in Fig. \ref{fig:mainFPC_truncated}a. The latter is fed to the PMPG framework, and the optimized $\mathbf{u_t}$ that minimizes $\mathcal{A}$ is obtained. As shown in Figs. \ref{fig:mainFPC_truncated}b and c, results demonstrate excellent agreement with the prediction of the CFD model which was obtained using a nontruncated domain of 15 $D$. This finding highlights a significant advantage of the PMPG framework: its ability to eliminate the dependence on predefined outflow boundary conditions. By enabling the use of small computational domains, the method significantly reduces computational costs while maintaining accuracy. This feature is particularly advantageous for investigating complex unsteady or periodic flows, where traditional approaches are limited by the constraints of finite domain size.
\textbf{}

\section{Temporal Evolution in PMPG}
The previous analysis focused on finding the optimal (i.e., dynamically correct) $\mathbf{u_t}$ at some time step with a given velocity field (i.e., initial condition) $\mathbf{u}(\mathbf{x},t)$. Once $\mathbf{u}_t$ is determined for this initial time step, it can be used to march the velocity field forward in time using any reasonable numerical integration scheme. As such, the flow field $\mathbf{u}(\mathbf{x},t+ \text{dt})$ at the next time step can be computed, which is then used in the cost (\ref{actionS}) to obtain the minimizing $\mathbf{u_t}$ at that time step, and so on. As such, in the PMPG formulation, time-marching is achieved without the chronic issues of pressure-velocity coupling; i.e. there is no need to solve Poisson equation in pressure at every time step.

\begin{figure}[!tbh]
\centering
\includegraphics[width=\textwidth]{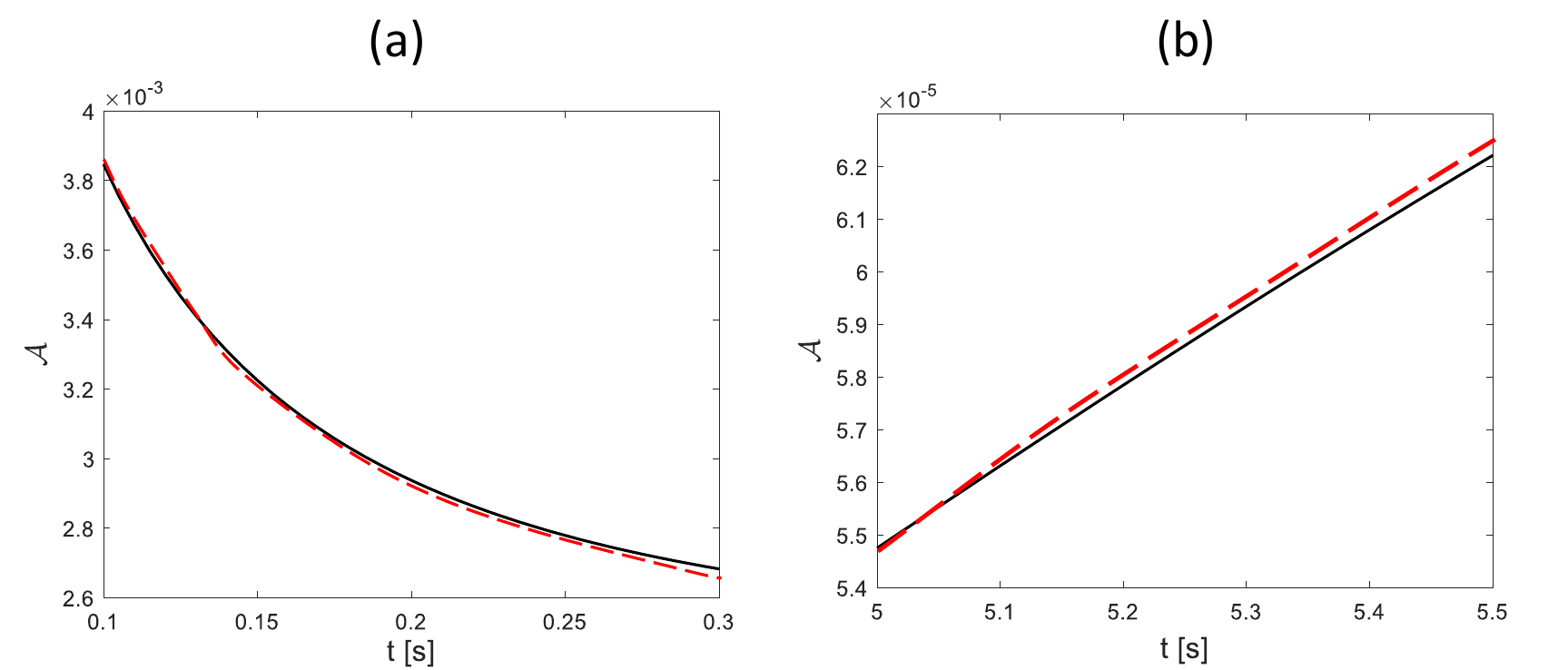}
\caption{Evolution of $\mathcal{A}$ with time for (a) $Re$ = 25, and (b) $Re$ = 500. The solid line represents high-fidelity CFD simulations, while the dashed lines represent results obtained using the unsteady PMPG-PINN framework}\label{fig:marchS}.
\end{figure}

We apply the time-stepping framework to the lid-driven cavity problem, focusing on two cases: $Re = 25$ and $Re = 500$. These cases are chosen specifically because the variation of $\mathcal{A}$ over time exhibits fundamentally different behavior. As shown in Fig.~\ref{fig:marchS}, at the lower Reynolds number, $\mathcal{A}$ decreases over time, whereas at the higher Reynolds number, $\mathcal{A}$ increases over time.

\begin{figure}[!tbh]
\centering
\includegraphics[width=0.50\textwidth]{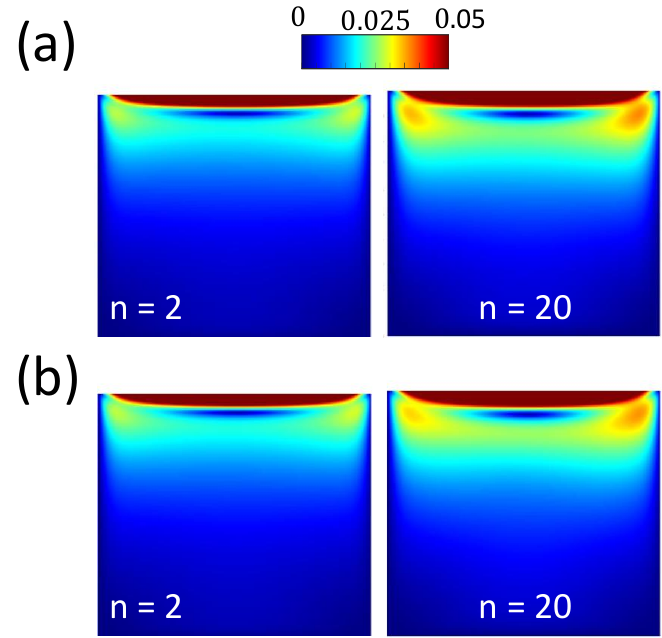}
\caption{(a) Predicted velocity field using PMPG-PINN. (b) Predicted velocity field using CFD simulations. Results are obtained for $Re = 25$ at $n=2$ and $n=20$.}\label{march_flow}.
\end{figure}

To march in time, we utilize the explicit Euler method \cite{atkinson1991introduction} for simplicity, although more accurate integration techniques could be applied. The update is performed using the following expression:
\begin{equation}
    \mathbf{u}^{n+1} = \mathbf{u}^{n} + \text{dt} \, \mathbf{u}^{n}_t,
\end{equation}
where $\text{dt}$ is chosen to maintain a Courant number below 1, following standard practices in high-fidelity simulations \cite{lomax2002fundamentals}. Simulations were performed over the time range $0.1 \le t \le 0.3$ $s$ for \textit{Re} = 25 and $5 \le t \le 5.5$ $s$ for \textit{Re} 500. The results shown in Fig.~\ref{fig:marchS} indicate that the PMPG-PINN framework, represented by the dashed line, accurately captures the evolution of the cost $\mathcal{A}$, whether it is increasing or decreasing. To further illustrate the accuracy of the temporal evolution using PMPG-PINN, Fig. \ref{march_flow} presents a comparison between the predicted flow field and CFD simulations for $Re = 25$ at two distinct time instants.

\begin{figure}[!tbh]
\centering
\includegraphics[width=\textwidth]{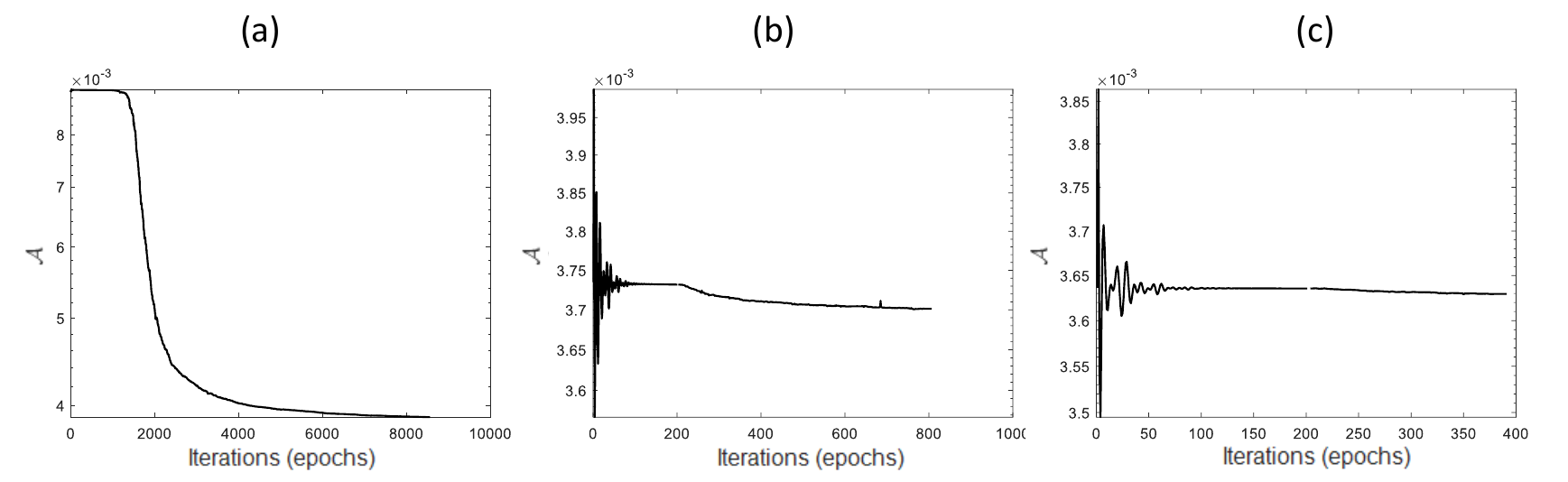}
\caption{Convergence of $\mathcal{A}$ at consecutive time steps during the temporal evolution of the PMPG-PINN. (a) $n = 1$, (b) $n = 2$, and (c) $n = 3$.}\label{marchResid}.
\end{figure}

Finally, we assess the performance of the algorithm in solving transient problems. Figure \ref{marchResid} displays the number of iterations required for convergence as time progresses (i.e., as the number $n$ of time steps increases). Figure ~\ref{marchResid}(a) shows that convergence of the initial minimization at $n=1$ requires a relatively high number of iterations, reaching 8000 epochs. However, as shown in Fig. ~\ref{marchResid}(b), at the subsequent time steps ($n=2$), the number of iterations necessary for convergence decreases significantly to fewer than 800. This efficiency improvement is achieved through transfer learning \cite{weiss2016survey}, where previously acquired knowledge accelerates the convergence process in later stages. By the third time step ($n=2$), the number of iterations necessary for convergence further reduces to fewer than 400.

\section{Conclusions}
In this study, we developed a variational computational-based formulation for solving unsteady incompressible fluid mechanics problems, integrating the Principle of Minimum Pressure Gradient (PMPG) with Physics-Informed Neural Networks (PINNs). 

The method was validated through three classical benchmarks-the lid-driven cavity, Poiseuille flow, and the flow past a circular cylinder-demonstrating excellent agreement with high-fidelity CFD simulations, even in cases where conventional PINN methods struggle, such as high Reynolds number flows in lid-driven cavity.\\

The key advantages of the presented variational approach are as follows: 
\begin{itemize}
    \item By circumventing the pressure-velocity coupling, this method eliminates the need for solving the Poisson equation for pressure at each time step, significantly improving computational efficiency. As such, this method offers a promising alternative to traditional CFD approaches, particularly in large-scale simulations.
    
    \item This method overcomes limitations resulting from the need to impose nonphysical  boundary conditions at the outflow to solve CFD problems. Specifically, unlike traditional CFD methods, which require domain extension to prevent outflow distortions, this formulation can handle cases where the domain is finite, as demonstrated in the flow past a circular cylinder. 
    
    \item This method has a more rapid convergence when compared to conventional PINNs. This is primarily due to the elimination of pressure from the equations, which reduces the solution space, and the computational burden of auto-differentiation. 
\end{itemize}

Additionally, temporal evolution analysis revealed that this method can effectively generate transient solutions over time. While the initial time step may require more computational resources, subsequent iterations are considerably faster, highlighting the improving efficiency of the approach as time simulation progresses.

\newpage
\appendix
\setcounter{equation}{0}
\setcounter{figure}{0}
\setcounter{section}{0}
\appendix
\setcounter{equation}{0}
\setcounter{figure}{0}
\setcounter{section}{0}

\section{High Fidelity Simulations}

Computational fluid dynamics (CFD) involves numerically solving the governing equations that describe the behavior of a viscous fluid. In this work, these  equations are simplified to the incompressible Navier-Stokes equations as:
 \begin{eqnarray}
\nabla \cdot \mathbf{u} &= 0, 
\end{eqnarray}
and the momentum conservation can be written as
\begin{eqnarray}
\frac{\partial \mathbf{u}}{\partial t} + \mathbf{u} \cdot   \nabla \mathbf{u} - \nabla \cdot (\nu \nabla \textbf{u})  &= -\nabla {p},
\end{eqnarray}
 where $\mathbf{u} = (u, v)$ is the velocity vector field, $\displaystyle p$ is the kinematic pressure, and $t$ is time. 

To solve the governing equations, these are discretized into a system of algebraic equations using the finite volume method using the open-source CFD toolbox OpenFOAM\textregistered \footnote{OpenFOAM\textregistered \ adopts  finite-volume discretization techniques to achieve solutions for continuum mechanics problems using C++ library packages.  A detailed description of the package can be found in Ref.\cite{jasak1996error}.}. This method applies the conservation laws to control volumes (cells) within the computational domain, solving the equations in an Eulerian framework. This section details the computational domain, boundary conditions, spatial discretization, and the discretization schemes used in the CFD model.

\subsection{Computational Domain and Boundary Conditions}
The computational domain for the three examples is illustrated in Fig. \ref{schematic}, with four boundary patches: inlet, outlet, sides, and walls. For the Poiseuille flow, the domain extends to \( L = 2 \) in the \( x \)-direction and \( L = 1 \) in the \( y \)-direction. The lid-driven cavity flow is modeled in a square domain with \( L = 1 \) for both the \( x \)- and \( y \)-directions. In the case of flow past a circular cylinder, the domain size is carefully chosen to balance computational efficiency and accuracy. The inflow length is set to \( 10D \), and the overall cross-flow domain length is \( 20D \), where \( D \) is the diameter of the cylinder. An outflow length of \( L = 15D \) is used, ensuring the outflow boundary effects on the solution are negligible.

Table \ref{tab:BCs} depicts the boundary conditions applied in the model.
\begin{table}[]
\centering
\caption{Boundary conditions}
\label{tab:BCs}
\begin{tabular}{@{}ccccc@{}}
\toprule
Variable     & Inlet                                        & Outlet  & Sides    & Walls                                        \\ \midrule
$p$          & $\frac{\partial p}{\partial \mathbf{n}} = 0$ & $p$ = 0 & Symmetry & $\frac{\partial p}{\partial \mathbf{n}} = 0$ \\
$\mathbf{u}$ & $\mathbf{u} = Uin$                           & $\frac{\partial \mathbf{u}}{\partial \mathbf{n}} = 0$         & Symmetry         & $\mathbf{u} = 0$                                             \\ \bottomrule
\end{tabular}

\end{table}
A Neumann boundary condition was applied for the pressure at the inlet and at the wall, and for velocity at the outlet. The velocity at the cylinder wall, at the inlet, and the pressure at the outlet were specified with a Dirichlet boundary condition. The velocity at the cylinder wall was set to zero in all directions to ensure the no-slip condition. The symmetry boundary condition in OpenFOAM is interpreted as a wall patch with slip condition. This means the flow at the far field sides is in parallel with the uniform inlet flow. This is a safe modeling assumption since the upper and lower height of the computational domain is sufficiently large.

\subsection{Spatial Discretization and Numerical Schemes}
For spatial discretization, a structured mesh was used for the Poisueilli and lid-cavity flow. However, for flow past a circular cylinder, an unstructured mesh was used. The block around the cylinder was based on an O-grid mesh, see Ref.\cite{derksen2019numerical}. The regions around the moving body and the wake region are refined finely in order to capture the large fluctuations expected from the flow. The remaining regions in the domain has a coarsened grid to reduce the computational time.

The two-dimensional fluid equations are solved using the incompressible transient solver \textit{PimpleFoam} \cite{lindblad2014implementation}. The gradient, the divergence, and the Laplacian terms in the fluid equations are discretized using a second-order Gauss integration scheme. A first-order implicit Euler scheme is adopted to solve the transient terms. Finally, and before the results were utilized, a convergence study is performed to ensure that the numerical solution is independent of the grid size.

 \bibliographystyle{elsarticle-num}\biboptions{sort&compress}
\newpage
\bibliography{mybib}   

\end{document}